\documentclass[aps,pra,reprint,10pt,superscriptaddress,showpacs,floatfix]{revtex4-1}

\usepackage{amsmath}
\usepackage{graphicx}
\usepackage{float}

\begin{document}

\title{Approach for making visible and stable stripes in a spin-orbit-coupled Bose-Einstein superfluid}

\author{Giovanni I. Martone}

\affiliation{INO-CNR BEC Center and Dipartimento di Fisica, Universit\`a di Trento, I-38123 Povo, Italy}

\author{Yun Li}

\affiliation{INO-CNR BEC Center and Dipartimento di Fisica, Universit\`a di Trento, I-38123 Povo, Italy}
\affiliation{Centre for Quantum Technologies, National University of Singapore, 3 Science Drive 2, Singapore 117542}

\author{Sandro Stringari}

\affiliation{INO-CNR BEC Center and Dipartimento di Fisica, Universit\`a di Trento, I-38123 Povo, Italy}

\begin{abstract}
The striped phase exhibited by a spin-$1/2$ Bose-Einstein condensate with spin-orbit coupling is characterized by the spontaneous breaking of two continuous symmetries: gauge and translational symmetry. This is a peculiar feature of supersolids and is the consequence of interaction effects. We propose an approach to produce striped configurations with high-contrast fringes, making their experimental detection in atomic gases a realistic perspective. Our approach, whose efficiency is directly confirmed by three-dimensional Gross-Pitaevskii simulations, is based on the space separation of the two spin components into a two-dimensional bilayer configuration, causing the reduction of the effective interspecies interaction and the increase of the stability of the striped phase. We also explore the effect of a $\pi/2$ Bragg pulse, causing the increase of the fringe wavelength, and of a $\pi/2$ rf pulse, revealing the coherent nature of the order parameter in the spin channel.
\end{abstract}


\maketitle

The experimental realization of synthetic gauge fields and spin-orbit-coupled configurations in spinor Bose-Einstein condensates  \cite{Lin2009_PRL,Lin2009_Nature,Lin2011_NatPhy,Lin2011_Nature} has opened interesting perspectives for the realization of novel quantum phases. Among them the striped phase represents one of the most challenging configurations, being characterized by the spontaneous breaking of two continuous symmetries: gauge and translational symmetry. The simultaneous breaking of these symmetries is a typical feature of supersolids, a phase of matter not yet realized experimentally, despite systematic efforts made in solid $^4$He \cite{Balibar2010}. The realization of supersolidity is presently the object of several theoretical proposals, focusing on atomic gases interacting with dipolar \cite{Goral2002,Sansone2010,Pollet2010} or soft-core, finite-range forces \cite{Cinti2010,Saccani2011,Saccani2012,Kunimi2012,Macri2012,Cinti2014,Gross}.

The occurrence of a striped phase in spin-orbit-coupled two-component Bose gases has been the object of several recent theoretical investigations  \cite{Wang2010,Wu2011,Ho2011,Sinha2011,Ozawa2012,Li2013,Zezyulin2013,Lan2014,Sun2014,Han2014,Hickey2014}. The experiment of  Ref. \cite{Lin2011_Nature} was not, however, able to reveal the typical features of this phase, characterized by the occurrence of periodic modulations of the density profile. The main reason is that the contrast and the wavelength of the modulations in the experimental conditions of \cite{Lin2011_Nature} are too small. When written in a locally spin-rotated frame \cite{Martone2012}, the single-particle spin-orbit Hamiltonian realized in \cite{Lin2011_Nature} takes the following form:
\begin{equation}
h_{\rm sp}= \frac{1}{2m}[\left(p_x - \hbar k_0 \sigma_z\right)^2 + p_\perp^2] + \frac{\hbar\Omega}{2} \, \sigma_x + \frac{\hbar\delta}{2} \, \sigma_z \, .
\label{Hsp}
\end{equation}
This Hamiltonian is the result of the application of two counterpropagating polarized lasers with wave vector difference ${\bf k}_0$, chosen along the $x$ direction, providing Raman transitions between two different hyperfine states, in the presence of a nonlinear Zeeman field. The strength of the Raman coupling is fixed by the parameter $\Omega$. Equation (\ref{Hsp}) also includes the coupling with an effective magnetic field $\delta$ given by the sum of the true external magnetic field and of the frequency detuning between the two lasers (see, for example, \cite{Martone2012}). The spin matrices entering the single-particle Hamiltonian are the usual $2\times2$ Pauli matrices. The operator ${\bf p}=-i\hbar \nabla$ is the canonical momentum, the physical velocity being given by $({\bf p} \mp \hbar k_0 \hat{\bf e}_x)/m$ for spin-up and spin-down particles, respectively, with $\hat{\bf e}_x$ the unit vector along the $x$ direction. A peculiar feature of the Hamiltonian $h_{\rm sp}$, when $\delta=0$, is the occurrence of a degenerate double minimum in momentum space capable of hosting a condensate with canonical momentum  $p_x = \pm \hbar k_1 = \pm \hbar k_0 \sqrt{1 - (\hbar\Omega/4 E_r)^2}$, $E_r=(\hbar k_0)^2/2m$ being the recoil energy. The wave vector $k_1$ differs from $k_0$ if $\Omega \ne 0$ and vanishes for $\hbar\Omega=4 E_r$. For larger values of $\Omega$ the gas is in the single-minimum phase where all the atoms occupy the ${\bf  p}=0$ single-particle state. Another peculiar feature caused by the Raman term is that, while the total number of atoms is conserved, the population of the single spin states depends crucially on the parameters of the Hamiltonian and cannot be fixed {\it a priori}.  By adding a position dependence in the magnetic field, and hence in the parameter $\delta$,  it is possible to generate effective gauge fields which simulate the presence of a Lorentz force. This procedure has been employed to generate quantized vortices in Bose-Einstein condensates \cite{Lin2009_Nature}. Even in the absence of space dependence in the magnetic field the phase diagram associated with the spin-orbit Hamiltonian (\ref{Hsp}) is very rich. The quantum phases emerging in these spin-orbit-coupled gases actually depend in a crucial way on the interactions, for which we use the mean-field expression
\begin{equation}
H_{\rm int} = \int {\rm d}^3r \left[\frac{g_{\uparrow\uparrow}}{2} n_\uparrow({\bf r})^2 + \frac{g_ {\downarrow\downarrow}}{2} n_\downarrow({\bf r})^2 + g_{\uparrow\downarrow} n_\uparrow({\bf r}) n_\downarrow({\bf r})\right] ,
\label{Hint}
\end{equation}
where $g_{\alpha\beta}=4\pi\hbar^2 a_{\alpha\beta}/m$ ($\alpha,\beta=\,\uparrow,\downarrow$) are the coupling constants in the different spin channels, fixed by the corresponding scattering lengths $a_{\alpha\beta}$, while $n_{\uparrow,\downarrow}$ are the densities of the two spin components.
In this Rapid Communication we consider spin configurations characterized by the condition $g_{\uparrow\uparrow}g_{\downarrow\downarrow} > g_{\uparrow\downarrow}^2$, which ensures miscibility between the two spin components in the absence of Raman coupling and of external magnetic fields. The combined presence of the spin-orbit term proportional to $k_0$ and of the Raman coupling $\Omega$ can give rise to a demixed configuration, also called plane-wave phase, where all the atoms occupy a single-particle state with canonical momentum $p_x=\hbar k_1$, or $p_x=-\hbar k_1$. 
In the weak coupling limit $\bar{n} g_{\alpha\beta} \ll E_r$, where $\bar{n}$ is the average density, the transition between the mixed and the demixed phase is predicted \cite{Ho2011,Li2012} to occur at the density-independent value
\begin{equation}
\hbar \Omega_{\rm tr}(\gamma)= 4 E_r \sqrt{\frac{2\gamma}{1+2\gamma}} \, ,
\label{Omegatr}
\end{equation}
where $\gamma=G_2/G_1$  with $G_1= \bar{n}(g_{\uparrow\uparrow}+g_{\downarrow\downarrow}+2g_{\uparrow\downarrow})/8$ and $G_2 = \bar{n}(g_{\uparrow\uparrow}+g_{\downarrow\downarrow}-2g_{\uparrow\downarrow})/8$, and where we have taken the value $\hbar\delta=-\bar{n}(g_{\uparrow\uparrow}-g_{\downarrow\downarrow})/2$ in order to compensate the asymmetry caused by the difference between the spin up-up and spin down-down coupling constants \cite{Li2012}. The value of $\Omega_{\rm tr}$ turns out to be relatively small in systems with almost equal coupling constants. For example, in the case of the states $\left|\uparrow\right\rangle = \left| F=1,m_F=0\right\rangle$ and $\left|\downarrow\right\rangle = \left| F=1,m_F=-1\right\rangle$ of $^{87}$Rb,  where  $a_{\uparrow\uparrow}=101.41\,a_B$ and $a_{\downarrow\downarrow}=a_{\uparrow\downarrow}=100.94\,a_B$, $a_B$ being the Bohr radius, one finds $\hbar\Omega_{\rm tr} = 0.19 \,E_r$. The transition between a mixed and a demixed phase has been experimentally identified in \cite{Lin2011_Nature} and \cite{Ji2014} at the value of $\Omega_{\rm tr}$ predicted by theory.

The mixed phase  exhibits  density modulations in the form of stripes according to the law
\begin{equation}
n({\bf r})= \bar{n} \left[1 + \frac{\hbar\Omega}{2(2 E_r + G_1)}\cos(2k_1x+\phi) \right] ,
\label{density}
\end{equation}
whose contrast $(n_{\rm max}-n_{\rm min})/(n_{\rm max}+n_{\rm min})$ is given by the ratio $\hbar\Omega/[2(2 E_r+G_1)]$, and whose periodicity is determined by the wavelength $\pi/k_1$, with $k_1 = k_0 \sqrt{1-\{\hbar\Omega/[2(2 E_r+G_1)]\}^2}$. The occurrence of stripes has been predicted in \cite{Li2013} to give rise to a characteristic double gapless band structure in the excitation spectrum $\varepsilon({\bf q})$ as a function of the momentum transfer ${\bf q}$ along the direction orthogonal to the stripes, which could be measured using Bragg spectroscopy techniques. Result (\ref{density}) has been obtained assuming the simplified form
\begin{equation}
\Psi({\bf r}) = \sqrt{\frac{\bar{n}}{2}}\left[\begin{pmatrix} \cos \theta \\ -\sin\theta \end{pmatrix}
e^{i k_1x} + e^{-i\phi} \begin{pmatrix} \sin \theta \\ -\cos\theta
\end{pmatrix} e^{-i k_1x} \right]
\label{spinor}
\end{equation}
for the order parameter, with $2\theta = \arccos(k_1/k_0)$, which ignores higher order harmonics with wave vectors $\pm 3k_1$, $\pm 5k_1$, etc. \cite{Li2013}. The same order parameter gives a vanishing value for the spin density ($n_\uparrow=n_\downarrow$). This choice for the order parameter implies the breaking of translational invariance in the density profile, despite the translational invariance of the Hamiltonian, the actual position of fringes being fixed by the value of the phase $\phi$. In the plane-wave phase only one of the two momentum components of the order parameter (\ref{spinor}) is present, and the density profile is uniform.  Both the contrast and the periodicity of the density fringes in the striped phase are too small to be revealed in the case of the parameters used in \cite{Lin2011_Nature}, the predicted maximum contrast, reached at $\Omega=\Omega_{\rm tr}$, being equal to $\sim 0.04$. On the other hand, the sole observation of mixing ($n_\uparrow=n_\downarrow$) is not a proof of the coherent nature characterizing the linear combination (\ref{spinor}) of the order parameter. One should also point out that the difference $\Delta\mu$ between the chemical potentials in the demixed and the mixed phases is very  small. A simple estimate is obtained at $\Omega=0$ in the case $g_{\uparrow\uparrow}=g_{\downarrow\downarrow}$, where one finds  $\Delta\mu =  2 G_2$. The value of $\Delta \mu$ is even smaller at finite $\Omega$. As a consequence a tiny magnetic field (arising, for example, from external fluctuations) can easily bring the system into the spin-polarized demixed phase, unless one increases significantly the value of $G_2$.

The purpose of this work is to propose a strategy to produce striped configurations characterized by high-contrast and long-wavelength fringes, and to emphasize the coherent role of their order parameter in the spin channel. The same strategy yields a huge increase of the stability of the striped phase.

An efficient way to increase the contrast is to reduce the value of the crossed coupling constant $g_{\uparrow\downarrow}$, yielding an increase of $\gamma$ and hence of $\Omega_{\rm tr}$ [see Eq.~(\ref{Omegatr})]. A possibility is to look for hyperfine states characterized by a small (or tunable) interspecies scattering length. In this work we follow a different strategy, based on the trapping of the atomic gas in a two-dimensional (2D) configuration, with tight confinement of the spin-up and spin-down components  around two positions displaced by a distance $d$ along the $z$ direction. This configuration can be realized with a spin-dependent potential of the form  
\begin{equation}
V_\mathrm{ext}(z) =  \frac{m\omega^2_z}{2}\left(z-\frac{d}{2}\sigma_z\right)^2 \; ,
\label{Vextd}
\end{equation}
produced either through magnetic gradient techniques or via spin-dependent optical potentials. In the absence of Raman coupling the integration over $z$ of the interaction energy functional (\ref{Hint}) gives rise to  effective 2D coupling constants $\tilde{g}_{\alpha\beta}$ given by
\begin{equation}
\tilde{g}_{\uparrow\uparrow,\downarrow\downarrow}=\frac{1}{\sqrt{2\pi}a_z} g_{\uparrow\uparrow,\downarrow\downarrow} \; , \quad
\tilde{g}_{\uparrow\downarrow}=\frac{1}{\sqrt{2\pi}a_z} g_{\uparrow\downarrow}e^{-d^2/2a^2_z} \; ,
\label{gtilde}
\end{equation}
where we have used the Gaussian profile $\psi_\pm=(1/\sqrt[4]{\pi a_z^2})e^{-(z\mp d/2)^2/2a^2_z}$ for the $z$ dependence of the spin-up and spin-down wave functions, with $a_z=\sqrt{\hbar/m\omega_z}$ the oscillator length along the $z$ direction. Equation (\ref{gtilde}) explicitly shows that the effect of the relative displacement of the two densities causes a quenching of the interspecies coupling constant with respect to the two other components, and hence an enhancement of the ratio
\begin{equation}
\gamma = \frac{\tilde{G}_2}{\tilde{G}_1} = \frac{g_{\uparrow\uparrow}+g_{\downarrow\downarrow}-2g_{\uparrow\downarrow}e^{-d^2/2a^2_z}}{g_{\uparrow\uparrow}+g_{\downarrow\downarrow}+2g_{\uparrow\downarrow}e^{-d^2/2a^2_z}} \, .
\label{gamma}
\end{equation}
In an analogous way one finds that also the effective Raman coupling, to be used in 2D, is lowered with respect to the physical coupling $\Omega$ according to the law $\tilde{\Omega} =  e^{-d^2/4a^2_z}\Omega$,  reflecting the reduction of the overlap between the two wave functions.

In conclusion, in the presence of a spin-dependent  displacement caused by a tight axial  trapping potential, the new  configuration can be described formulating the Hamiltonian in 2D, with the effective Raman coupling given by $\tilde{\Omega}$ and the interaction term obtained from the functional (\ref{Hint}), with the replacement of the three-dimensional (3D) densities with their 2D counterparts $\int {\rm d}z \, n$ and of the coupling constants with the renormalized values (\ref{gtilde}). The main difference with respect to the original 3D problem is the increase of the ratio (\ref{gamma}) fixed by the value of $d$, with the consequent increase of the critical value of the Raman coupling and of the reachable contrast of fringes in the striped phase. For example, choosing the value $d=a_z$ and the $^{87}$Rb hyperfine states mentioned above, one finds the value $\gamma=0.25$ for the ratio (\ref{gamma}), to be compared with the value $\gamma =0.0012$ for the $d=0$ case. As a consequence the maximum reachable contrast takes much larger values \cite{Note1}.

\begin{figure}[t]
\centering
\includegraphics[scale=1]{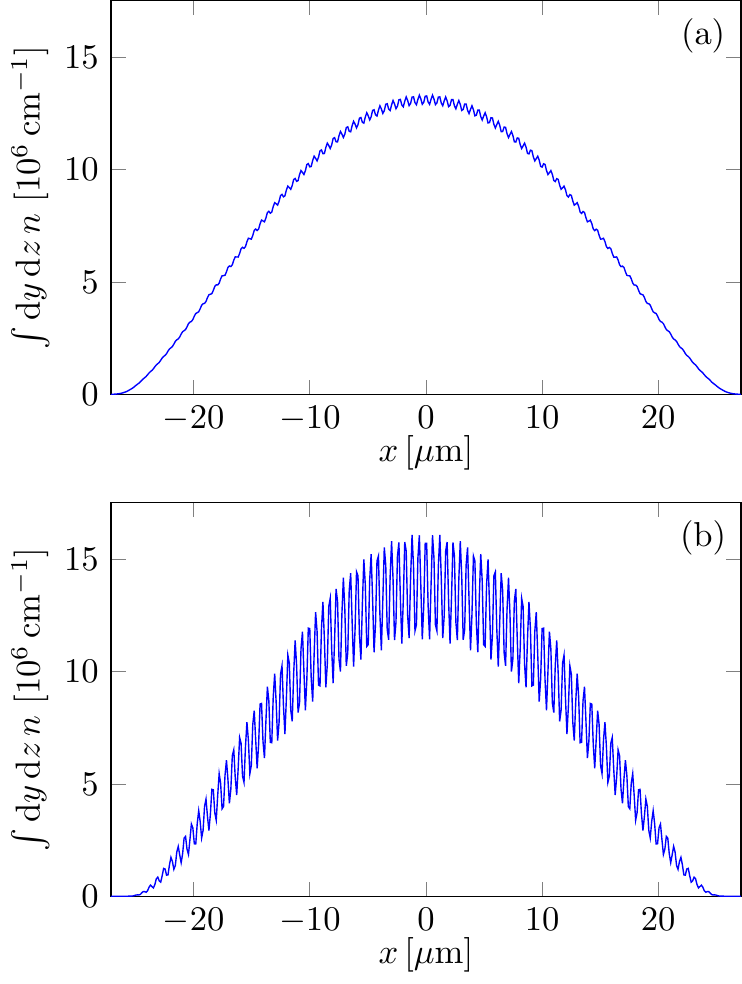}
\caption{(Color online) Integrated density profile $\int {\rm d}y \, {\rm d}z \, n$ in the striped phase. (a) displays the situation without separation of the traps for the two spin components. (b) corresponds instead to traps separated along $z$ by a distance $d=a_z$, which  increases the contrast of the fringes.}
\label{fig:density_prof}
\end{figure}

In order to check the validity of the 2D picture described above, and to provide quantitative predictions in real configurations, we have solved numerically the 3D Gross-Pitaevskii equation for a gas of $N=4\times 10^4$ rubidium atoms trapped by a 3D harmonic potential. The parameters $k_0=5.54\, \mu$m$^{-1}$ and $E_r = h\times 1.77\,$kHz are chosen consistently with Ref.~\cite{Lin2011_Nature}. The results are shown in Fig.~\ref{fig:density_prof} for the trapping frequencies $\left(\omega_x,\omega_y,\omega_z\right)=2\pi\times\left(25,100,2500\right)\,$Hz. Figure~\ref{fig:density_prof}(a)  corresponds to  $d=0$, while  Fig.~ \ref{fig:density_prof}(b) corresponds to $d=a_z=0.22\, \mu$m. In both Figs.~\ref{fig:density_prof}(a) and Fig.~\ref{fig:density_prof}(b) we have chosen values of the Raman coupling equal to one-half the critical value needed to enter the plane-wave phase, in order to ensure a larger stability to the stripe phase. In Fig.~\ref{fig:density_prof}(a) this corresponds to  $\hbar\Omega = (1/2) \hbar\Omega_{\rm tr}(\gamma)=0.095\, E_r$ ($\gamma=0.0012$) while in Fig.~ \ref{fig:density_prof}(b) to $\hbar\Omega = (1/2)e^{d^2/4a^2_z } \hbar\Omega_{\rm tr}(\gamma)=1.47\, E_r$  ($\gamma=0.25$). The plotted density corresponds to the one-dimensional (1D) density as a function of the most relevant $x$ variable, obtained by integrating the full 3D density along the $y$ and $z$ directions. The figure clearly shows that in the conditions of almost equal coupling constants [Fig.~\ref{fig:density_prof}(a)] the density modulations are very small, while their effect is strongly amplified in Fig.~\ref{fig:density_prof}(b) where the spin up-down coupling is quenched with respect to the spin up-up and spin down-down values by the factor $\sim 0.61$. We have also verified that, with the above choice of the parameters, the solution of the 2D Gross-Pitaevskii equations, with the same radial trapping conditions and the renormalized values $\tilde{g}_{\uparrow\downarrow}$ and $\tilde{\Omega}$, is not only in qualitative, but also quantitative agreement with the results of the full 3D Gross-Pitaevskii calculation reported in Fig.~\ref{fig:density_prof}(b). It is also worth noticing that, since the suggested procedure reduces significantly the value of the spin up-down constant and at the same time increases the value of the local 3D density, it also has the positive effect of significantly increasing the energy difference between the striped and the plane-wave phase, thereby making the former much more robust against magnetic perturbations.  For example, in the case considered in the above 3D Gross-Pitaevskii simulation with $d=a_z$ [Fig.~\ref{fig:density_prof}(b)], a magnetic detuning of the order of $0.37 \, E_r$ is needed to bring the system into the spin-polarized phase, while in the absence of displacement  [Fig.~\ref{fig:density_prof}(a)] the critical value is much smaller ($\sim 0.001 \, E_r$). We have also checked that the quality of stripes is not significantly affected using a relatively softer confinement along $z$. For example, using $\omega_z=2\pi \times 625\,$Hz, i.e., a factor 4 smaller, $d=a_z$ and the same value of $\Omega$, we find that the contrast is still significant ($0.14$ instead of $0.16$). The critical magnetic detuning needed to destabilize the stripe phase is reduced because of the smaller value of the local 3D density ($0.23 \, E_r$ instead of $0.37 \, E_r$).

\begin{figure}[t]
\centering
\includegraphics[scale=1]{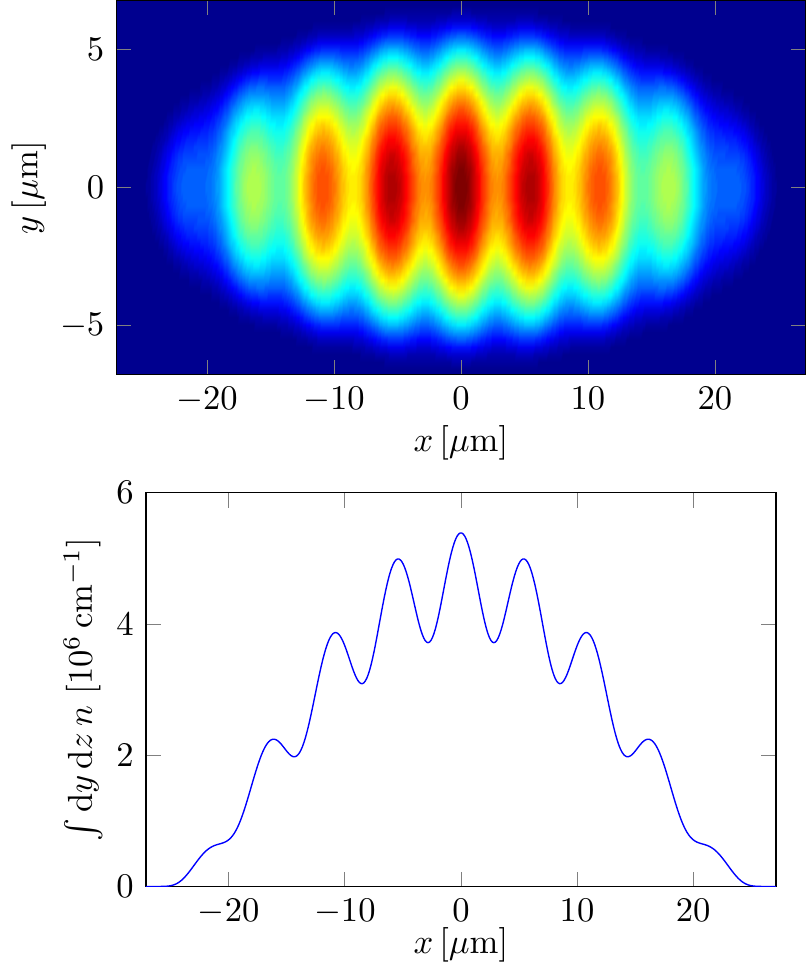}
\caption{(Color online) Integrated density profiles $\int {\rm d}z \, n$ (top) and $\int {\rm d}y \, {\rm d}z \, n$ (bottom) in the stripe phase, in the same conditions as Fig.~\ref{fig:density_prof}(b), after the application of a $\pi/2$ Bragg pulse transferring momentum $\pm 1.8\,\hbar k_1$.}
\label{fig:density_prof_Bragg}
\end{figure}

A major problem for the {\it in situ} identification of the  fringes is also caused by the smallness of their spatial separation, fixed by $\pi/k_1$, which turns out to be of the order of a fraction of a micron in standard conditions. In order to increase the stripe wavelength one possibility is to lower the value of $k_0$ by using lasers with a smaller relative incident angle. In the following we propose a more drastic and efficient procedure which consists of producing, after the realization of the striped phase,  a $\pi/2$ Bragg pulse with a short time duration (smaller than the time $\hbar/E_r$ fixed by the recoil energy), followed by the sudden release of the trap. This pulse can transfer to the condensate a momentum $p_B$ or $-p_B$ along the $x$ direction,  with $p_B=\hbar k_B$ chosen equal to $\hbar(2k_1 -\epsilon)$ with $\epsilon$ small compared to $k_1$. The $\pi/2$ pulse has the effect of splitting the condensate into various pieces, with different momenta. After the Bragg pulse two of these pieces will move slowly with momenta $\mp\hbar(k_0-k_1)$ and $\mp\hbar(k_0-k_1+\epsilon)$, respectively, where the minus (plus) sign refers to the spin-up (down) component, and will be able to interfere giving rise to interference fringes with wavelength $2\pi/\epsilon$, which can easily become large and visible {\it in situ}. The other pieces produced by the Bragg pulse carry much higher momenta and will fly away rapidly after the release of the trap and of the laser fields. It is worth noticing that the interference originates from the two momentum components (\ref{spinor}) of the order parameter and involves $1/3$ of the total number of atoms. It would be absent in the plane-wave phase. In order to observe such fringes the time of flight $\tau$ should be short enough to avoid separation between the interfering pieces which move with a slightly different velocity ($\tau < m \Delta X/\hbar\epsilon$, where $\Delta X$ is the size of the condensate in the $x$ direction). In Fig.~\ref{fig:density_prof_Bragg} we show a typical behavior of the density profile obtained by modifying the condensate wave function in momentum space according to the prescriptions discussed above, and choosing $\epsilon= 0.2\, k_1$.

\begin{figure}[!htb]
\centering
\includegraphics[scale=1]{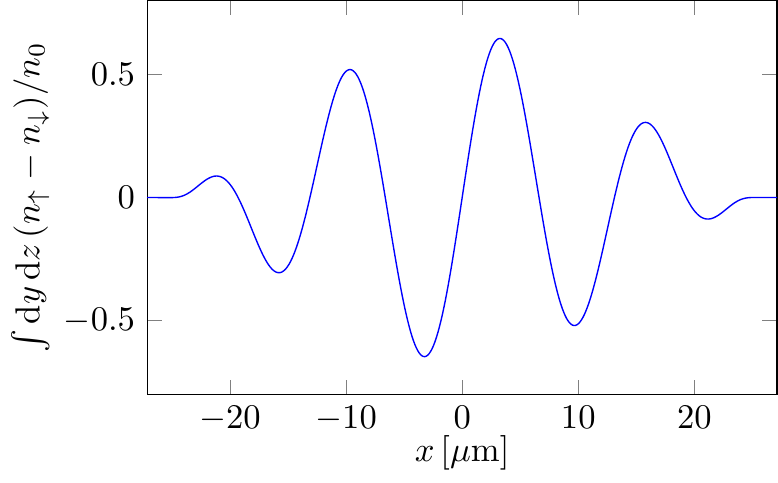}
\caption{(Color online) 3D Gross-Pitaevskii simulation for the integrated spin density profiles $\int {\rm d}y \, {\rm d}z \, (n_\uparrow-n_\downarrow)$ in the stripe phase, in the same conditions as Fig.~\ref{fig:density_prof}(b), after the application of a fast $\pi/2$ rf pulse.}
\label{fig:density_prof_rf}
\end{figure}

The coherent nature characterizing the two momentum components of the order parameter (\ref{spinor}) could be revealed by the application of a fast $\pi/2$ rf pulse described by the unitary transformation $\hat{U}=e^{i\theta_{\rm rf}\sigma_x/2}$ with $\theta_{\rm rf}=\pi/2$. The rf pulse mixes the two spin components of the order parameter and gives rise to interference fringes of wavelength $\pi/(k_0-k_1)$ in the spin density distribution $n_\uparrow-n_\downarrow$, as a consequence of the transformation law $\hat{U}^{-1}\sigma_z\hat{U}= \cos\theta_{\rm rf} \, \sigma_z - \sin\theta_{\rm rf} \, \sigma_y$ and of the resulting interference effect associated with the spin average of the transverse operator $\sigma_y$ in the striped phase. The spin density, after the $\pi/2$ rf pulse, takes the form 
\begin{equation}
\frac{n_\uparrow-n_\downarrow}{\bar{n}} = \frac{k_0 + k_1}{2k_0} \sin[2(k_0-k_1)x]
\label{fringesspin}
\end{equation}
apart from an unimportant phase factor, plus additional rapidly oscillating terms associated with higher momentum components. The total density is instead unaffected by the rf pulse. In Fig.~\ref{fig:density_prof_rf} we show the results of our 3D Gross-Pitaevskii simulation, where $n_0$ is the 1D total density
calculated in the center of the trap in the absence of spin-orbit coupling. In the figure we have included only the long wavelength modulations in the calculation of the spin density $n_\uparrow-n_\downarrow$.

In conclusion we have proposed a combined procedure to increase the contrast of fringes and their wavelength, thereby favoring the visibility of the  density modulations characterizing the stripe phase in realistic experimental conditions. The suggested procedure has the important effect of producing a sizable increase of  the energetic stability of the striped phase, protecting it from  fluctuations of the magnetic field. The 2D geometry proposed in this Rapid Communication would also open interesting perspectives in the study of the Berezinskii-Kosterlitz-Thouless phase transition in 2D Bose superfluids containing stripes.

We would like to thank Jean Dalibard, Gabriele Ferrari, Giacomo Lamporesi, Lev P. Pitaevskii, and Ian Spielman for stimulating discussions. This work was supported by the ERC through the QGBE grant and by Provincia Autonoma di Trento.


\begin{thebibliography}{99}
\bibitem{Lin2009_PRL} Y.-J. Lin, R. L. Compton, A. R. Perry, W. D. Phillips, J. V. Porto, and I. B. Spielman, Phys. Rev. Lett. {\bf 102}, 130401 (2009).
\bibitem{Lin2009_Nature} Y.-J. Lin, R. L. Compton, K. Jim\'{e}nez-Garc\'{i}a, J. V. Porto, and I. B. Spielman, Nature (London) {\bf 462}, 628 (2009).
\bibitem{Lin2011_NatPhy} Y.-J. Lin, R. L. Compton, K. Jim\'{e}nez-Garc\'{i}a, W. D. Phillips, J. V. Porto, and I. B. Spielman, Nat. Phys. {\bf 7}, 531 (2011).
\bibitem{Lin2011_Nature} Y.-J. Lin, K. Jim\'{e}nez-Garc\'{i}a, and I. B. Spielman, Nature (London) {\bf 471}, 83 (2011).
\bibitem{Balibar2010} S. Balibar, Nature (London) {\bf 464}, 176 (2010).
\bibitem{Goral2002} K. G\'{o}ral, L. Santos, and M. Lewenstein, Phys. Rev. Lett. {\bf 88}, 170406 (2002).
\bibitem{Sansone2010} B. Capogrosso-Sansone, C. Trefzger, M. Lewenstein, P. Zoller, and G. Pupillo, Phys. Rev. Lett. {\bf 104}, 125301 (2010).
\bibitem{Pollet2010} L. Pollet, J. D. Picon, H. P. B\"uchler, and M. Troyer, Phys. Rev. Lett. {\bf 104}, 125302 (2010).
\bibitem{Cinti2010} F. Cinti, P. Jain, M. Boninsegni, A. Micheli, P. Zoller, and G. Pupillo, Phys. Rev. Lett. {\bf 105}, 135301 (2010).
\bibitem{Saccani2011} S. Saccani, S. Moroni, and M. Boninsegni, Phys. Rev. B {\bf 83}, 092506 (2011).
\bibitem{Saccani2012} S. Saccani, S. Moroni, and M. Boninsegni, Phys. Rev. Lett. {\bf 108}, 175301 (2012).
\bibitem{Kunimi2012} M. Kunimi and Y. Kato, Phys. Rev. B {\bf 86}, 060510(R) (2012).
\bibitem{Macri2012} T. Macr\`{i}, F. Maucher, F. Cinti, and T. Pohl, Phys. Rev. A {\bf 87}, 061602(R) (2013).
\bibitem{Cinti2014} F. Cinti, T. Macr\`{i}, W. Lechner, G. Pupillo, and T. Pohl, Nat. Commun. {\bf 5}, 3235 (2014).
\bibitem{Gross} The occurrence of density modulations in the ground state of a dilute Bose gas interacting with soft-core, finite-range interactions was first predicted by Gross \cite{Gross57}.
\bibitem{Gross57} E. P. Gross, Phys. Rev. {\bf 106}, 161 (1957).
\bibitem{Wang2010} C. Wang, C. Gao, C.-M. Jian, and H. Zhai, Phys. Rev. Lett. {\bf 105}, 160403 (2010).
\bibitem{Wu2011} C.-J. Wu, I. Mondragon-Shem, and X.-F. Zhou, Chin. Phys. Lett. \textbf{28}, 097102 (2011).
\bibitem{Ho2011} T.-L. Ho and S. Zhang, Phys. Rev. Lett. {\bf 107}, 150403 (2011).
\bibitem{Sinha2011} S. Sinha, R. Nath, and L. Santos, Phys. Rev. Lett. {\bf 107}, 270401 (2011).
\bibitem{Ozawa2012} T. Ozawa and G. Baym, Phys. Rev. A \textbf{85}, 063623 (2012).
\bibitem{Li2013} Y. Li, G. I. Martone, L. P. Pitaevskii, and S. Stringari, Phys. Rev. Lett. {\bf 110}, 235302 (2013).
\bibitem{Zezyulin2013} D. A. Zezyulin, R. Driben, V. V. Konotop, and B. A. Malomed, Phys. Rev. A \textbf{88}, 013607 (2013).
\bibitem{Lan2014} Z. Lan and P. \"Ohberg, Phys. Rev. A \textbf{89}, 023630 (2014).
\bibitem{Sun2014} Q. Sun, L. Wen, W.-M. Liu, G. Juzeli{\= u}nas, and An-Chun Ji, arXiv:1403.4338.
\bibitem{Han2014} W. Han, G. Juzeli{\= u}nas, W. Zhang, and W.-M. Liu, arXiv:1407.2972.
\bibitem{Hickey2014} C. Hickey and A. Paramekanti, arXiv:1409.1216.
\bibitem{Martone2012} G. I. Martone, Y. Li, L. P. Pitaevskii, and S. Stringari, Phys. Rev. A {\bf 86}, 063621 (2012).
\bibitem{Li2012} Y. Li, L. P. Pitaevskii, and S. Stringari, Phys. Rev. Lett. {\bf 108}, 225301 (2012).
\bibitem{Ji2014} S.-C. Ji, J.-Y. Zhang, L. Zhang, Z.-D. Du, W. Zheng, Y.-J. Deng, H. Zhai, S. Chen, and J.-W. Pan, Nat. Phys. {\bf 10}, 314 (2014).
\bibitem{Note1} Another important consequence of the new spin bilayer configuration concerns the value of the critical density $n^{(c)}=2E_r/[\gamma(\tilde{g}_{\uparrow\uparrow}+\tilde{g}_{\downarrow\downarrow})]$, needed to reach the tricritical point \cite{Li2012} where the striped, the plane-wave, and the single-minimum phases meet. The value of $n^{(c)}$ is actually significantly reduced with respect to the corresponding 3D case, due to the much larger value of $\gamma$.
\end{thebibliography}
\end{document}